\documentclass[prb,rsi,reprint]{revtex4-2}
\usepackage[utf8]{inputenc}
\usepackage[T1]{fontenc}
\usepackage{graphicx}
\usepackage{hyperref}

\begin{document}
\title{Tuning the superconducting dome in granular aluminum thin films}
\author{Aniruddha Deshpande}
\affiliation{1.~Physikalisches Institut, Universit{\"a}t Stuttgart, D-70569 Stuttgart, Germany}

\author{Jan Pusskeiler}
\affiliation{1.~Physikalisches Institut, Universit{\"a}t Stuttgart, D-70569 Stuttgart, Germany}

\author{Christian Prange}
\affiliation{1.~Physikalisches Institut, Universit{\"a}t Stuttgart, D-70569 Stuttgart, Germany}

\author{Uwe Rogge}
\affiliation{1.~Physikalisches Institut, Universit{\"a}t Stuttgart, D-70569 Stuttgart, Germany}

\author{Martin Dressel}
\affiliation{1.~Physikalisches Institut, Universit{\"a}t Stuttgart, D-70569 Stuttgart, Germany}

\author{Marc Scheffler}
\affiliation{1.~Physikalisches Institut, Universit{\"a}t Stuttgart, D-70569 Stuttgart, Germany}

\date{\today}

\begin{abstract}

Granular aluminum, which consists of nanometer-sized aluminum grains separated by aluminum oxide, is a peculiar superconductor. Its phase diagram as function of normal-state resistivity features a superconducting dome with a maximum critical temperature $T_\textrm{c}$ well above the $T_\textrm{c} = 1.2~\textrm{K}$ of pure aluminum. 
Here we show how the maximum $T_\textrm{c}$ of this superconducting dome grows if the substrate temperature during deposition is lowered from 300~K to cooling with liquid nitrogen (150~K and 100~K) and liquid helium (25~K). The highest $T_\textrm{c}$ we observe is 3.27~K. 
These results highlight that granular aluminum is a model system for complex phase diagrams of superconductors and demonstrate its potential in the context of high kinetic inductance applications. 
This is augmented by our observation of comparably sharp superconducting transitions of high-resistivity samples grown at cryogenic temperatures and by a thickness dependence even for films substantially thicker than the grain size.
\begin{description}
\vspace{4mm}
\item[Keywords]
Superconductivity, granular aluminum, grain size, superconducting dome
\end{description}
\end{abstract}

\maketitle

\section{Introduction}

Strongly disordered as well as granular superconductors can have very low superfluid density $n_\textrm{s}$, i.e.\ the density of Cooper pairs in the superconducting condensate can be much smaller than in crystalline superconductors \cite{Dutta2022}. 
Such superconductors with low $n_\textrm{s}$ can host various unconventional electronic states including Cooper pair insulators or pseudogap states \cite{Sacepe2011}. 
These states often depend on the level of disorder in the material, which thus acts as a convenient tuning parameter during material growth. 
The corresponding phase diagrams, where the critical temperature $T_\textrm{c}$ of superconductivity typically is suppressed with increasing disorder, have been studied in great detail in the context of the superconductor-insulator transition 
\cite{Sacepe2020}.
Manifestations of low $n_\textrm{s}$ are long London penetration depth $\lambda_\textrm{L}$ (from a magnetic-field point of view) and high kinetic inductance $L_\textrm{kin}$ (from an electrodynamic point of view) \cite{Meservey1969}.
Superconductors with high $L_\textrm{kin}$ play an important role as microwave circuit elements \cite{Niepce2019}, or for parametric amplifiers \cite{Eom2012,Malnou2021,Zapata2024arXiv} within the context of superconducting quantum computing \cite{Winkel2020}.
Furthermore, strongly disordered superconductors are the core material for superconducting nanowire single photon detectors (SNSPDs)
 \cite{EsmaeilZadeh2021}.

\begin{figure}
\includegraphics[width=8.5cm]{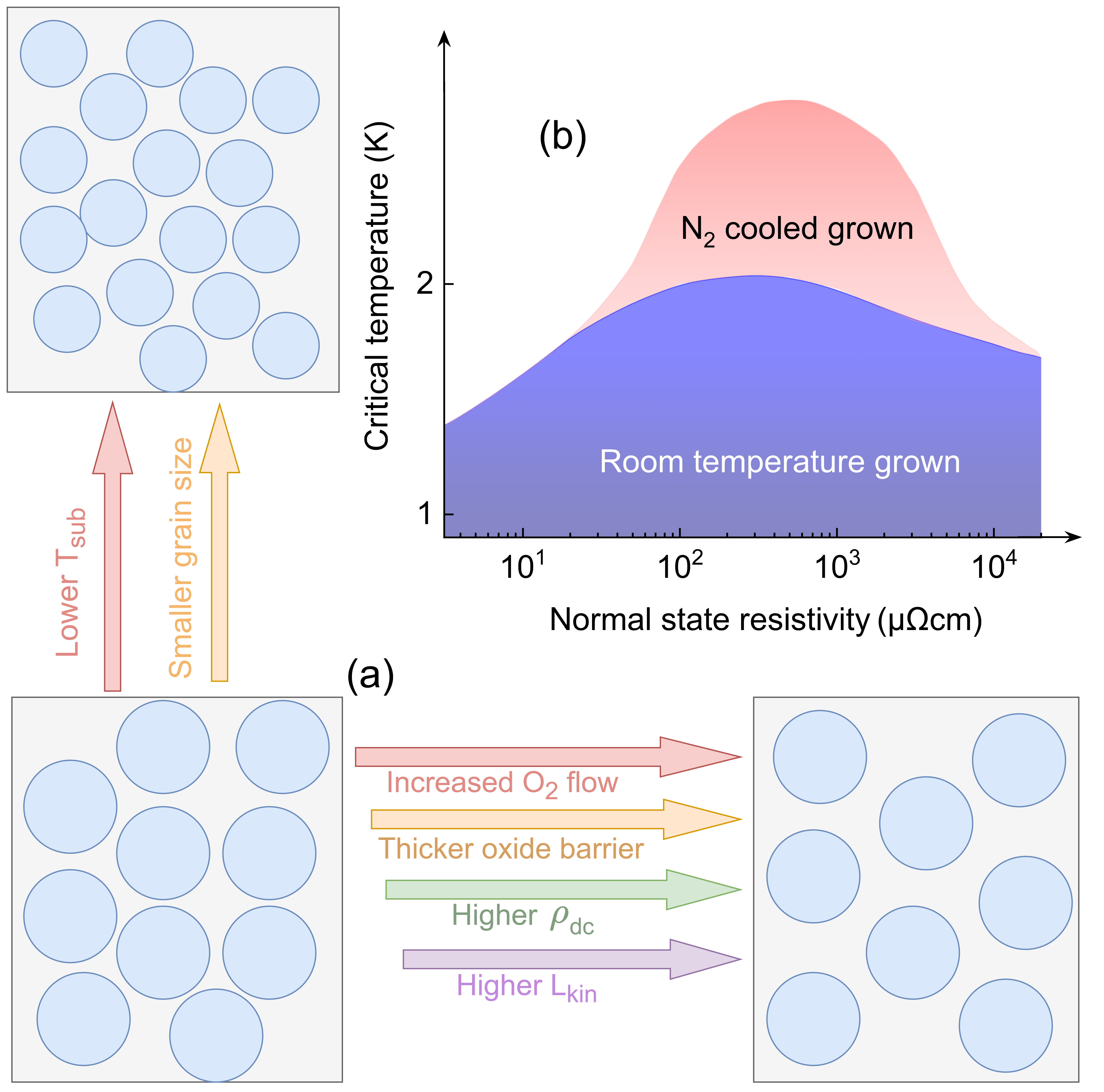}
\caption{ (a) Schematic superconductivity in granular aluminum: pure aluminum grains are separated by aluminum oxide, forming a distinct macroscopic superconductor. More oxygen inflow during growth increases dc resistivity, thus pushing towards the right of the phase diagram. Lower substrate temperature leads to smaller grains and higher critical temperature $T_\textrm{c}$. (b) Schematic superconducting dome in the phase diagram of granular aluminum, with enhanced $T_\textrm{c}$ and a higher dome for deposition on nitrogen-cooled substrates (compared to room-temperature substrates).}
\label{Fig:SchematicSuperconductivityInGrAl}
\end{figure}

Granular aluminum plays a prominent role among the wider class of disordered superconductors due to several characteristics \cite{Deutscher2021,Bachar2023}: 
(1) preparation of granular aluminum films is straightforward by deposition of aluminum vapor in an oxygen atmosphere (e.g.\ using thermal or electron-beam evaporation or sputtering) \cite{Deutscher1973-05,Rotzinger2016,Gruenhaupt2019,Kamenov2020}. 
(2) The normal-state resistivity and the superfluid density can be tuned over several orders of magnitude \cite{Deutscher1973-05,Pracht2016}. 
(3) With increasing normal-state resistivity, $T_\textrm{c}$ first increases substantially before then decreasing, leading to a phase diagram with a so-called superconducting dome \cite{Deutscher1973-05}. 
(4) These properties are advantageous in the field of quantum circuits and superconducting quantum computation, and therefore granular aluminum has been employed in numerous studies of concrete applications \cite{Gruenhaupt2018,Zhang2019,Moshe2020,Kamenov2020}.

Thus it might come as a surprise that the underlying mechanisms that control superconductivity in granular aluminum, including the shape of the superconducting dome, have been at the center of long-standing discussions. General considerations about material characteristics of granular aluminum are sketched in Fig.\ \ref{Fig:SchematicSuperconductivityInGrAl}(a), and the previously established resulting phase diagram with superconducting dome \cite{Deutscher1973-01,Deutscher1973-05,Pracht2016,Valenti2019,Moshe2019} is shown schematically in Fig.\ \ref{Fig:SchematicSuperconductivityInGrAl}(b).

Granular aluminum consists of nanometer-sized grains of pure aluminum within an aluminum oxide matrix. With overall properties being metallic, this means that neighbouring aluminum grains are strongly coupled via thin oxide barriers \cite{Cohen1968}. 
Microscopically, the superconducting state of granular aluminum thus can be considered a random three-dimensional array of superconducting islands coupled via Josephson junctions, while the macroscopic electronic properties of granular aluminum are similar to those of homogeneously disordered superconductors \cite{Steinberg2008,Pracht2016,Pracht2013,Moshe2019}. 
The normal-state dc resistivity $\rho_\textrm{dc}$ is a commonly used material property to quantify grain decoupling in granular aluminum, which in the phase diagram (see Fig.\ \ref{Fig:SchematicSuperconductivityInGrAl}(b)) is a parameter that can be tuned during growth. 
Previous studies have demonstrated that the initial increase of $T_\textrm{c}$ for increasing $\rho_\textrm{dc}$ is due to an increase of the superconducting energy gap $\Delta$ while the subsequent decrease of $T_\textrm{c}$ is caused by suppressed superfluid density $n_\textrm{s}$ \cite{Pracht2016,LevyBertrand2019}.
The size of the aluminum grains is another crucial parameter. 
Most experiments with superconducting granular aluminum employ thin films that were deposited on room-temperature substrates, where the diameter of the aluminum grains is around 3~nm, with maximum $T_\textrm{c}$ of the superconducting dome around 2.2~K \cite{Deutscher1973-01}. 
If instead the films are deposited onto substrates cooled with liquid nitrogen, then the grain size is around 2~nm, and maximum $T_\textrm{c}$ is around 3.2~K \cite{Deutscher1973-05,Moshe2021}. Superconductivity in such films grown at nitrogen temperatures was researched intensely by the Tel Aviv group \cite{Bachar2015-01,Moshe2019,Moshe2021,Deutscher2021}. Earlier related investigations into superconducting aluminum films utilized even colder substrates \cite{Buckel1954,Strongin1968,Sixl1974}, though with different scientific aims, with different fabrication, and with challenges concerning sample stability. 
Therefore, the goal of the present study is investigating the evolution of the superconducting dome of granular aluminum as a function of substrate temperature, from room temperature down to cooling with liquid helium.

\section{Sample preparation and measurement}

The granular aluminum films presented in this work were prepared by thermal evaporation of high-purity aluminum from a tungsten boat onto sapphire (Al$_2$O$_3$) substrates mounted on a copper substrate holder held at a distance of 28 cm from the boat. The substrate holder plate is directly coupled to the cryogenic tank and has built-in heater and sensor for temperature control. 
The films were grown in presence of a constant oxygen flow in the chamber to achieve the granular structure. An adjustable and gradually increasing current was passed through the tungsten boat to control the rate of evaporation of the aluminum. 
The substrate was kept at a constant temperature with the help of the heater. The thickness of the film was calculated using a quartz crystal installed inside the evaporation chamber that acts as rate meter.

The pressure in the deposition chamber before evaporation, in absence of oxygen, ranged between $~6\times 10^{-7}\, \textrm{mbar}$ (for substrates at 300~K) and $~1.5-2\times 10^{-7}\, \textrm{mbar}$ (for substrates at 25~K). 
To reach cold substrate temperatures, liquid nitrogen (N$_2$; boiling temperature of 77~K) or liquid helium (He; boiling temperature of 4~K) was introduced in the cryostat above the sample holder.

Two main process variables that affect the resistivity of the films are oxygen flow in the chamber and rate of evaporation of aluminum. The lower the oxygen flow, the lower is the resistivity due to reduced aluminum oxide barriers between the aluminum grains. 
Similarly, the higher the evaporation rate, the more aluminum is deposited on the substrate with respect to the incorporated oxygen, hence decreasing the resistivity. The films were prepared with resistivities ranging from 4 to 7000~$\mu\Omega$cm. 
To cover this range, the oxygen flow in the chamber was varied between 0.1 and 1.0~SCCM (standard cubic centimeters per minute). Similarly, the rate of evaporation of aluminum was varied between 0.8 and 3.5~\AA/s. The growth rate is set to a nearly constant level before the shutter in front of the substrate is opened.

For substrates cooled with liquid nitrogen, we investigated samples grown either at 150~K or 100~K substrate temperature. For helium-cooled substrates, the temperature was 25~K, which was the lowest temperature that we could routinely stabilize during the growth process of the films.

After film growth, the samples were taken out of the evaporator to ambient, and then contacted for temperature-dependent dc resistance measurements, which were performed in separate $^4$He-cooled bath cryostats.
Most of the samples were grown on $10 \times 10$~mm$^2$ substrates (suitable for THz measurements \cite{Bachar2015-04,Pracht2016,Moshe2019}) and studied using 4-point dc measurements in van der Pauw geometry (with contacts at the corners of the sample). 
A few samples were grown on $5 \times 5$~mm$^2$ substrates and then contacted using gold Corbino contacts (suitable for GHz measurements \cite{Scheffler2005}) deposited through a shadow mask in a separate thermal evaporator. Here dc resistance was measured in 2-point configuration, and the offset due to wiring was determined from a reference measurement.
For all samples, the dc resistivity was obtained from the dc resistance, the known lateral geometry, and the thickness as determined during growth with the quartz rate meter.
The normal-state resistivity at 5~K is used to relate the samples within the phase diagram below, and $T_\textrm{c}$ was defined as the temperature where the measured resistivity amounts to half of this normal-state value. 

\section{Results and discussion}

\begin{figure}
\centering
\includegraphics[width=8.5cm]{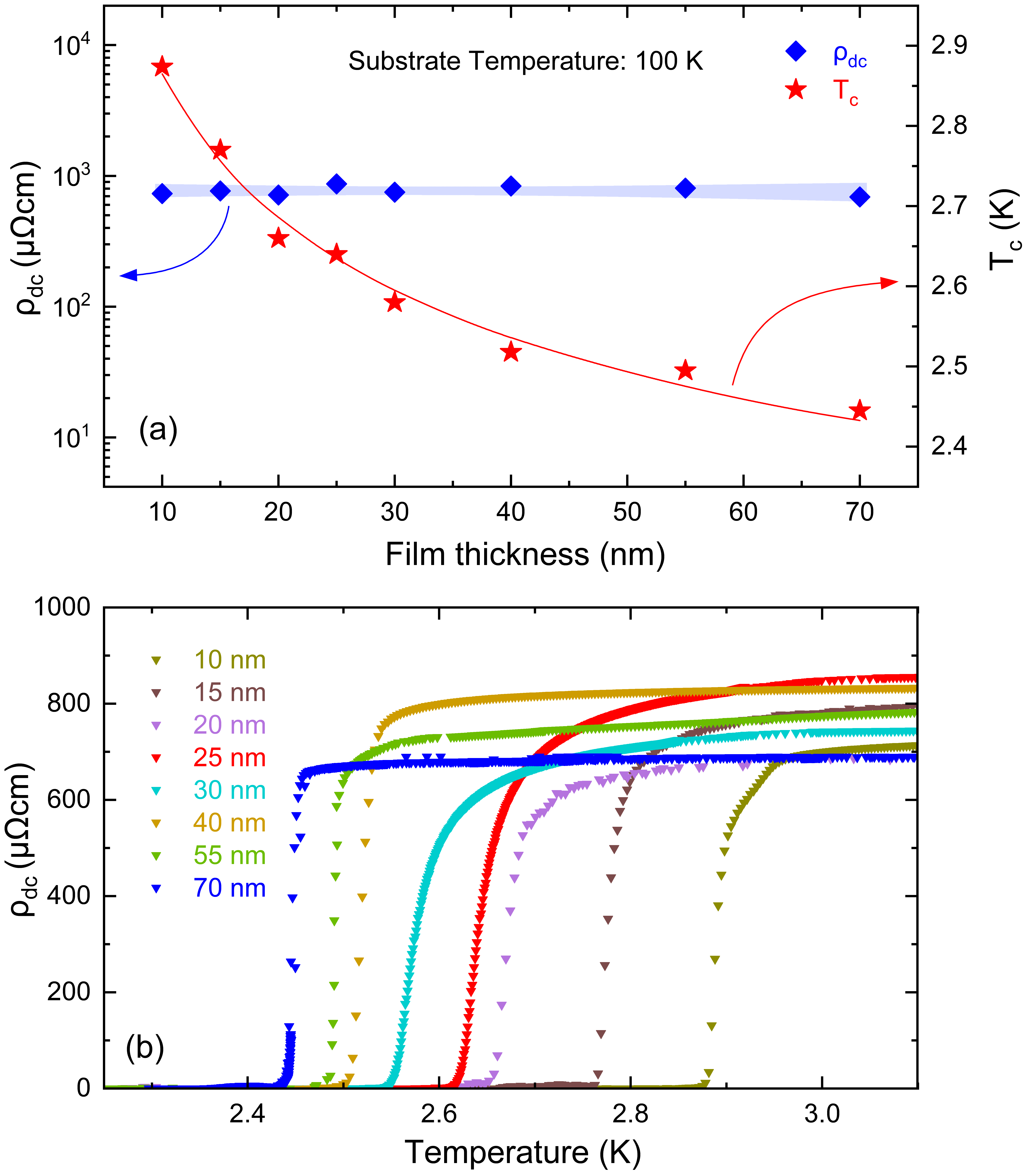}
\caption{Thickness dependence of critical temperature $T_\textrm{c}$ for superconducting granular aluminum films grown at 100~K. (a) Normal-state resistivity (left axis) and $T_\textrm{c}$ (right axis) as function of film thickness $d$. The red and shaded blue lines are guides to the eye.
(b) Temperature-dependent resistivity for these samples with their superconducting transitions.}
\label{Fig:ThicknessDependence}
\end{figure}

\subsection{Thickness dependence}

For superconducting films prepared from pure aluminum, previous studies have demonstrated a clear increase of $T_\textrm{c}$ above the bulk value $T_\textrm{c} = $~1.2~K if the thickness $d$ is reduced, in particular below 20~nm (with $T_\textrm{c} \approx $~1.4~K) and even more pronounced below 10~nm, while the actual numbers depend a lot on details of the sample fabrication  \cite{Chubov1969,Meservey1971,Arutyunov2019,Yeh2023}.

For granular aluminum films the grain size is a key length scale governing $T_\textrm{c}$ \cite{Abeles1966}, whereas the thickness is considered to be less important. Therefore, discussion of the superconducting dome as presented in Fig.\ \ref{Fig:SchematicSuperconductivityInGrAl} usually does not consider the sample thickness, which for many studies in this context is several tens of nanometers up to 100~nm.

In line with the current interest in superconducting films with preferably high kinetic inductance, we focus on granular aluminum films of thickness between 10~nm and 20~nm, several times the expected grain size of 3~nm or smaller. But during the course of this work, we realized an influence of the film thickness also for granular aluminum. 
Therefore, we grew several films at the same substrate temperature of 100~K, with film thickness $d$ ranging from 10~nm to 70~nm. 
The desired value $\rho_\textrm{dc} \approx 800 \, \mu\Omega\textrm{cm}$ was chosen on the top of the superconducting dome, where small variations of $\rho_\textrm{dc}$ have little impact on $T_\textrm{c}$. The superconducting transitions of these samples are shown as resistivity data in Fig.\ \ref{Fig:ThicknessDependence}(b) and the resulting values for $T_\textrm{c}$ in Fig.\ \ref{Fig:ThicknessDependence}(a).

For these films, $T_\textrm{c}$ clearly depends on film thickness, in a way that qualitatively resembles the behavior of pure aluminum films.
But while for the latter a simple $1/d$ behavior was found \cite{Meservey1971}, here we observe that the decrease of $T_\textrm{c}$ with increasing $d$ is weaker. This indicates that the thickness is less important here, which is consistent with the granular structure setting an intrinsic relevant length scale, namely the grain size of approximately 2~nm, which is still smaller but not much smaller than the investigated film thicknesses.

\begin{figure}
\includegraphics[width=8.7cm]{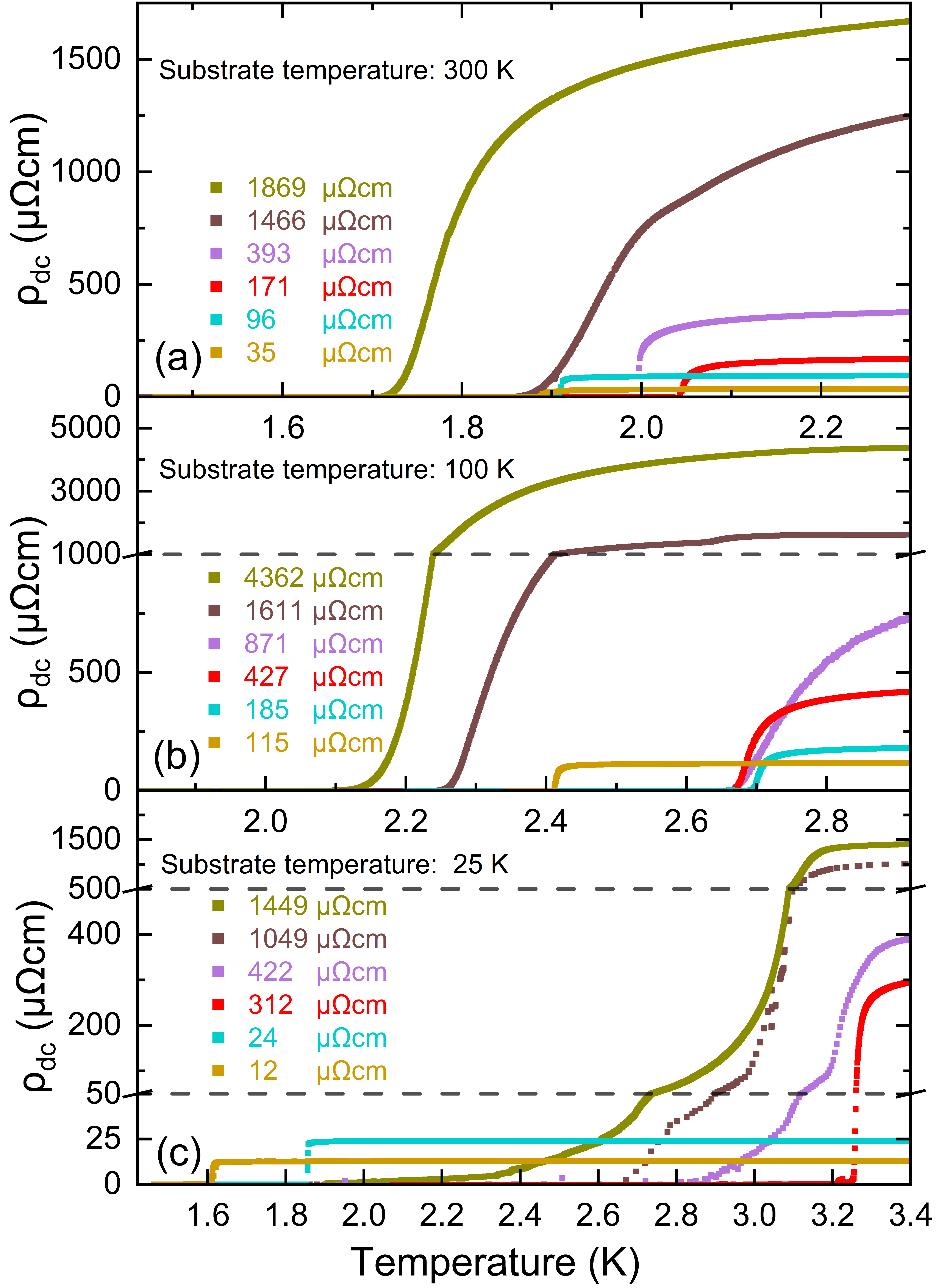}
\caption{Superconducting transitions for exemplary granular aluminum samples grown with substrate temperatures of (a) 300~K, (b) 100~K, and (c) 25~K. Dashed horizontal lines indicate scale changes for the vertical axes.}
\label{Fig:RhoDcTransitions}
\end{figure}

\subsection{Superconducting domes}

The resistive transitions for exemplary samples grown at three different substrate temperatures are shown in the three panels of Fig.\ \ref{Fig:RhoDcTransitions}. In total, these samples span more than three orders of magnitude in normal-state resistivity, and the corresponding $T_\textrm{c}$ varies substantially. 
These plots also indicate that the superconducting transitions in general become broader for the higher-resistivity samples, although one should not be misled due to the different vertical scale regions.

\begin{figure}
\centering
\includegraphics[width=8.7cm]{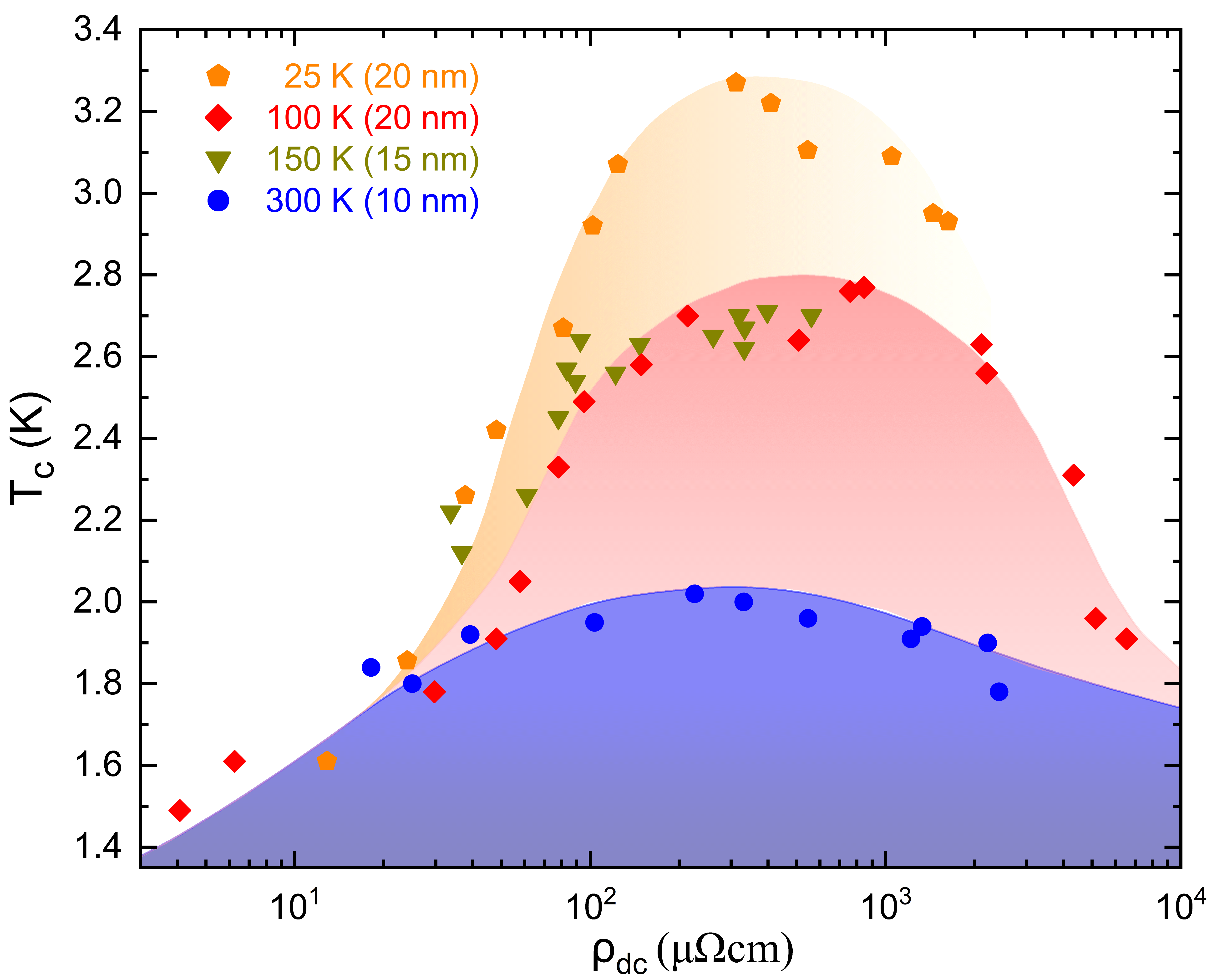}
\caption{Superconducting domes for granular aluminum grown at substrate temperatures between 25~K and 300~K. Film thickness (10~nm, 15~nm, or 20~nm) is the same amongst data for each of the substrate temperatures. Borders of shaded regions are guides to the eye.}
\label{Fig:SuperconductingDomes}
\end{figure}

The resulting phase diagrams, i.e.\ superconducting domes for different substrate temperatures, are shown in Fig.\ \ref{Fig:SuperconductingDomes}. The main result is that the highest $T_\textrm{c}$, i.e.\ the top of the superconducting dome, increases for reduced substrate temperature, and thus our highest $T_\textrm{c} = 3.27\, \textrm{K}$ is obtained for our lowest substrate temperature of 25~K. This confirms the trend previously observed when comparing samples grown at room temperature and with nitrogen cooling \cite{Deutscher1973-05}. (Additional data concerning substrate temperatures between 150~K and 290~K are presented in the Appendix.)

Here it might seem surprising that the superconducting domes that we find for substrate temperatures 150~K and 100~K basically lie on top of each other. But this can be explained by the different film thickness, with the 100~K-grown films of 20~nm being thicker than the 150~K-grown films of 15~nm. Thus, for these two data sets the $T_\textrm{c}$ enhancements due to lower substrate temperature (smaller grains) and lower film thickness compensate to coincidental overlap.

Furthermore, the maximum of the superconducting domes seem to be located in the same normal-state resistivity range, 300~$\mu\Omega$cm to 800~$\mu\Omega$cm, for all investigated substrate temperatures. This is consistent with previous studies of the superconducting dome of granular aluminum \cite{Deutscher1973-05, Dynes1984, Bachar2013, Moshe2019, LevyBertrand2019, Valenti2019}. 

The fundamental mechanism causing the strong enhancement of superconducting energy gap $\Delta$ and $T_\textrm{c}$ has been debated since its first observation \cite{Deutscher2021}. For the left side of the dome, the shell effect can explain the $T_\textrm{c}$ enhancement and highlights the role of the grain size \cite{GarciaGarcia2008,Bose2014,Pracht2016}, which can be tuned by substrate cooling.
For the right side of the dome it has been demonstrated experimentally that the superfluid density strongly decreases with increasing $\rho_\textrm{dc}$, and thus it is assumed that here $T_\textrm{c}$ is governed by the superfluid stiffness $J$, i.e.\ the energy scale that characterizes the resilience of the superfluid condensate to fluctuations of the phase of the condensate \cite{Emery1995}. 
For such superconductors in the dirty limit, one often estimates $J$ via $J \propto \Delta / \rho_\textrm{dc}$. 
Assuming that $\Delta$ is basically constant on the right side of the dome \cite{Pracht2016}, then $T_\textrm{c}$ should be proportional to $J$ and thus inversely proportional to $\rho_\textrm{dc}$. Comparing the behavior of $J$ as suggested by Pracht \textit{et al.} \cite{Pracht2016} with our observed $T_\textrm{c}$, we find the previously reported convergence of $T_\textrm{c}$ and $J$ for samples grown under nitrogen cooling. 
For samples grown at 25~K, the higher maximum $T_\textrm{c}$ suggests a larger energy gap due to smaller grain size and thus, following $J \propto \Delta / \rho_\textrm{dc}$, larger $J$ is expected for a given $\rho_\textrm{dc}$ (assuming that $\rho_\textrm{dc}$ is governed by the oxide between the grains and not by the pure aluminum grains). 
Thus the right flank of the \lq 25~K dome\rq{} could be shifted slightly upwards/to the right when compared to the \lq 100~K dome\rq. Unfortunately, the highest $\rho_\textrm{dc}$ that we have obtained for samples grown under helium cooling is only 1630~$\mu\Omega$cm, and thus too low for definite statements concerning the slope on the right side of the dome. Here, further studies towards even higher resistivity are desired.

On the left side of the phase diagram, our data would be consistent with all these investigated superconducting domes fanning out around $\rho_\textrm{dc} \approx 20 \, \mu\Omega\textrm{cm}$. 
Also here more detailed studies would be of interest to confirm whether there indeed is such a unique value of $\rho_\textrm{dc}$ that separates overlapping from differing behavior or whether the curves separate more gradually.

\subsection{Width of superconducting transitions}
\begin{figure}
\centering
\includegraphics[width=8.5cm]{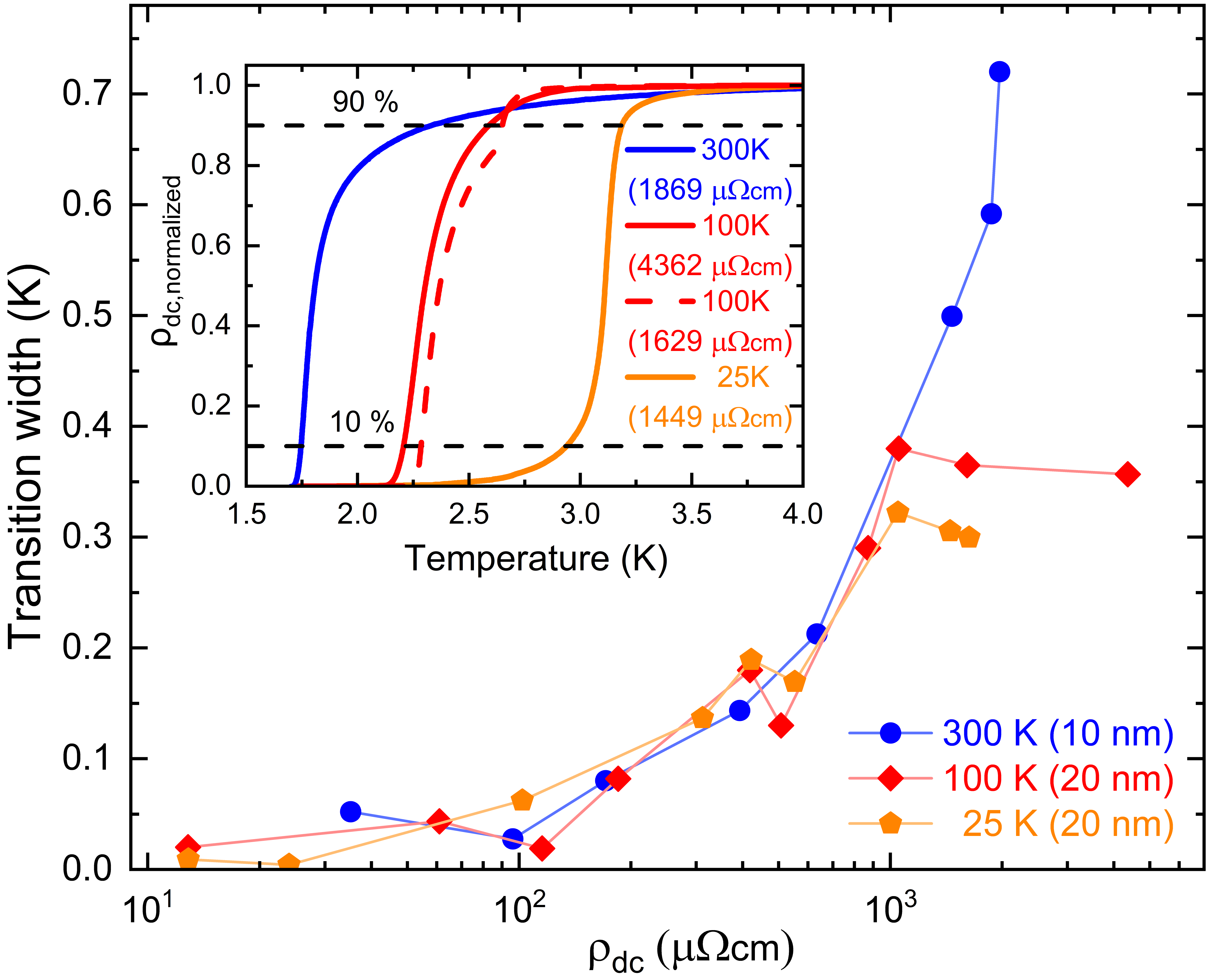}
\caption{Width of superconducting transition for samples grown at different substrate temperatures (and with thicknesses of 10~nm or 20~nm, respectively). The inset shows the normalized temperature-dependent resistivity for four samples with high $\rho_\textrm{dc}$. Here the dashed lines at 10\% and 90\% indicate the range that defines the transition width.
}
\label{Fig:TransitionWidth}
\end{figure}

For certain applications as well as for fundamental material characterization, the width of a superconducting transition is important. For granular aluminum, it has been observed that the superconducting transition becomes broader with higher $\rho_\textrm{dc}$ of the sample \cite{Deutscher1973-05}.
In Fig.\ \ref{Fig:TransitionWidth} we show the transition width, defined as the temperature difference between the two temperatures where the sample resistivity equals 90\% and 10\%, respectively, of the normal-state resistivity at 5~K, for samples grown at substrate temperatures 300~K, 100~K, and 25~K. Here we see the expected trend with low transition width for low $\rho_\textrm{dc}$ and a clear increase with increasing $\rho_\textrm{dc}$. Interestingly, for the samples grown at 100~K and 25~K, the transition width seems to not increase further for $\rho_\textrm{dc}$ beyond 1000~$\mu\Omega$cm. As seen in the inset of Fig.\ \ref{Fig:TransitionWidth} and in Fig.\ \ref{Fig:RhoDcTransitions}(a) for the high-$\rho_\textrm{dc}$ samples grown at 300~K there is prominent rounding of the curve at the higher-temperature edge of the superconducting transition, as is commonly discussed as Aslamazov-Larkin paraconductivity for homogeneously disordered superconductors and for granular aluminum \cite{Aslamazov1968,PrachtDissertation2017}. This effect is less pronounced for samples grown at lower temperatures. This might possibly be explained with previous observations \cite{BacharDissertation2017,Moshe2018,Deutscher2021} that the grain-size distribution for films grown at cryogenic temperatures is more homogeneous and thus that disorder effects are less relevant.

\section{Conclusions}
We have grown granular aluminum films at substrate temperatures ranging from room temperature down to 25~K. We find that the superconducting dome of $T_\textrm{c}$ becomes higher for decreasing substrate temperature, thus extending to the helium temperature range the trend previously observed down to nitrogen temperatures. 
This suggests that the reduction of grain size, which was previously characterized for room temperature and nitrogen temperature, continues into the helium temperature range. Clearly, structural characterization of films grown at helium temperatures is highly desired to confirm this evolution of grain size.

Often granular aluminum is considered to be a material system that is rather unsusceptible to the exact details of sample preparation, i.e.\ the type of disorder in the system generated by growth of aluminum grains with oxide barriers in between generally seems to be rather robust. 
Our study indicates that the story of superconductivity in granular aluminum might be even more intricate than already established. One aspect is the substantial thickness dependence. 
Another is the fact that the maximum of the superconducting domes that we observe for substrate temperatures of 300~K and 100~K is somewhat lower than previously reported in literature \cite{Deutscher1973-01,Deutscher1973-05,Pracht2016,Valenti2019,Moshe2019}. This might be important for quantitative optimization of devices that utilize the kinetic inductance of superconducting granular aluminum, since $T_\textrm{c}$ is not \lq universal\rq{} even for room-temperature growth.

This makes the picture of superconductivity in granular aluminum more complex, and it asks for more theoretical investigations of this fascinating model system for superconductivity with mesoscopic disorder. It also means that additional tuning parameters during growth might be possible, which could be helpful during material optimization for specific applications. In this context, our observation that the transition width of granular aluminum grown on cryogenically cooled substrates, for the region of high resistivity and thus high kinetic inductance, is narrower than for room-temperature growth, might be particularly attractive. It also reinforces the general interest in unconventional behavior of superconductors with low superfluid density, e.g.\ including collective excitations \cite{Buisson1994,Sherman2015,Pracht2017,Maleeva2018,LevyBertrand2019}, which now might become accessible with an even wider choice of tuning options.

\section*{Acknowledgments}
We thank Gabriele Untereiner for support with sample preparation, Alessandro D’Arnese for initial work on the evaporator, and Nimrod Bachar, Ameya Nambisan, and Ioan Pop for helpful discussions.
We acknowledge financial support by the Deutsche Forschungsgemeinschaft DFG (project SCHE 1580/6-1 / DR 228/57-1) and by the Baden-W\"urttemberg Stiftung within project QT-10 (QEDHiNet).

\begin{figure}
\centering
 \includegraphics[width=8.5cm]{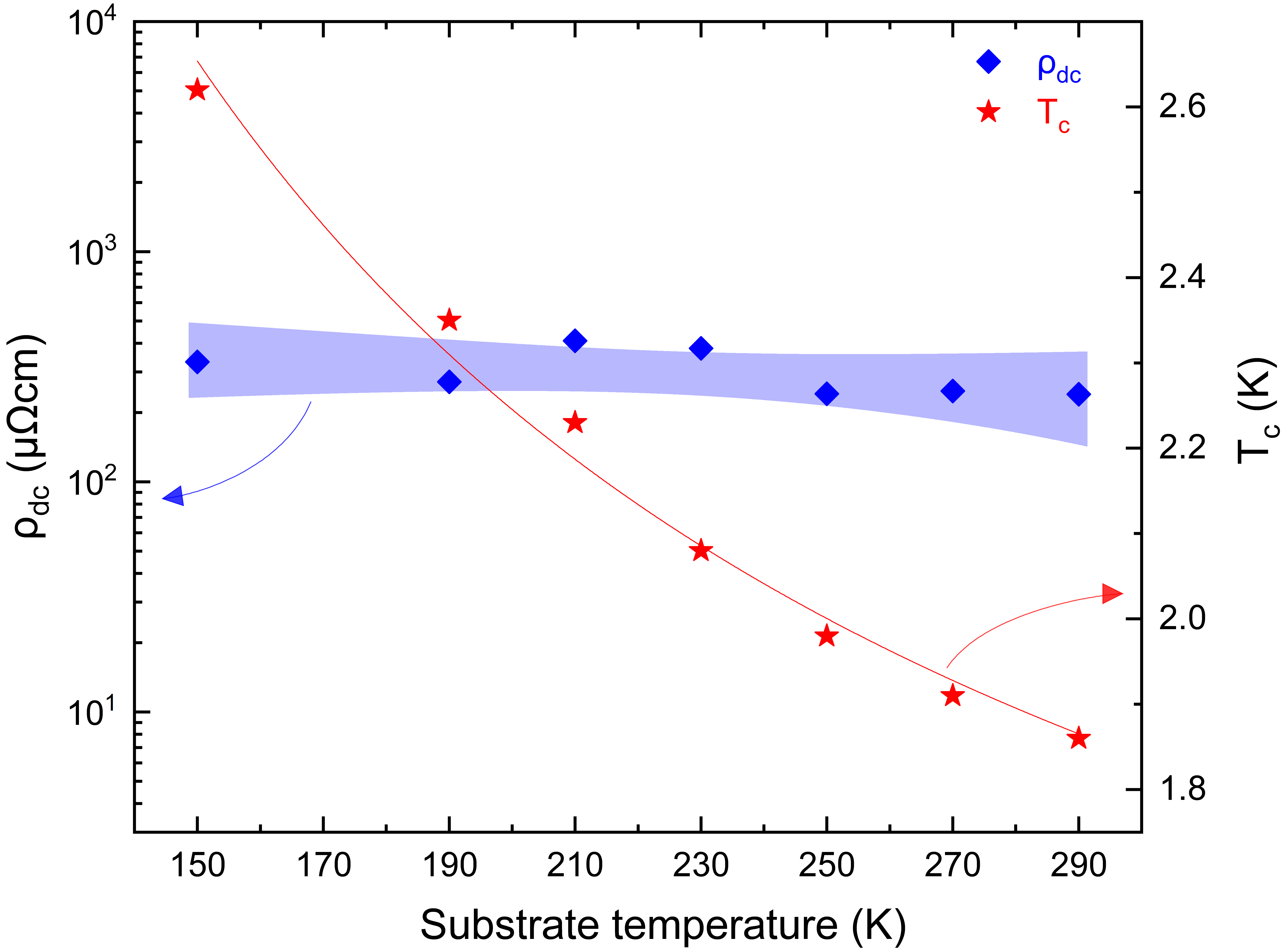}
\caption{Role of substrate temperature for superconducting transition. 10~nm thick samples 
were grown with similar normal-state resistivity $\rho_\textrm{dc}$ (left axis) at varying substrate temperature. The critical temperature $T_\textrm{c}$ continuously increases with decreasing substrate temperature.
Red line and blue-shaded range are guides to the eye.
}
\label{Fig:AppendixSubstrateTemperature}
\end{figure}
 
\section*{Appendix: continuous evolution of dome height}
We have prepared a set of films matching closely the top of the \lq 150~K\rq -dome, with thickness 10~nm and $\rho_\textrm{dc} \approx 300 \mu\Omega \textrm{cm}$, while changing the substrate temperature between 150~K and 290~K. The resulting values for $T_\textrm{c}$ are shown in Fig.\ \ref{Fig:AppendixSubstrateTemperature}.
We find a continuous reduction of $T_\textrm{c}$ with increasing substrate temperature. This confirms the general conception that grain size decreases continuously for cooler substrates and that this induces higher $T_\textrm{c}$. Thus any cooling of the substrate, even if it does not reach the \lq nitrogen range\rq{} around 77~K to 100~K, can significantly enhance $T_\textrm{c}$.

\bibliography{granularAl_2024-08-27} 

\end{document}